\documentclass[12pt,reqno,twoside]{amsart}%
\usepackage{amsmath,amsthm,amscd,amsthm,upref,indentfirst}
\usepackage{amsfonts,mathrsfs}

\usepackage{amssymb,amsbsy,bm}
\usepackage{graphicx}
\usepackage{times}
\usepackage{amsaddr}

\usepackage{soul}
\usepackage{amsxtra}
\usepackage{latexsym}
\usepackage{mathrsfs}
\usepackage{amsthm,upref,indentfirst}
\usepackage{amsbsy}
\usepackage{mathdots}
\usepackage{mathabx}

\usepackage{mathabx}

\usepackage[usenames,dvipsnames]{color}

\theoremstyle{plain}
\newtheorem{theorem}{Theorem}

\theoremstyle{definition}
\newtheorem{definition}[theorem]{Definition}
\newtheorem{criterion}[theorem]{Innovation}

\theoremstyle{remark}

\newtheorem{example}[theorem]{Example}
\numberwithin{equation}{section}
\numberwithin{theorem}{section}

\usepackage{exscale}
\usepackage[scaled=0.86]{helvet}

\normalfont\upshape

\usepackage[scr=euler,scrscaled=1.1,bb=txof,bbscaled=1.16,cal=boondox,calscaled=1.1]{mathalfa}
\renewcommand{\mathit}{\bm}
\renewcommand{\mathfrak}{\mathscr}

\renewcommand{\mathtt}[1]{\scalebox{1.2}{\bf \texttt{\upshape#1}}}
\renewcommand{\emph}[1]{\textcolor{blue}{\textbf{#1}}}

\usepackage{float}
\usepackage[justification=centering]{caption}

\numberwithin{equation}{section}
\numberwithin{theorem}{section}

\usepackage[normalem]{ulem}
\usepackage[square,comma]{natbib}

\usepackage{mparhack}

\usepackage{keyval}
\usepackage{totcount}
\newtotcounter{citnum} %From the package documentation
\def\oldbibitem{} \let\oldbibitem=\bibitem
\def\bibitem{\stepcounter{citnum}\oldbibitem}

\usepackage[verbose]{backref}
\backrefsetup{verbose=false}
\renewcommand*{\backref}[1]{}
\renewcommand*{\backrefalt}[4]{[{\tiny%
    \ifcase #1 \textsl{Not cited}%
          \or \textsl{Cited on page}~\textcolor{BrickRed}{#2}%
          \else \textsl{Cited on pages}~\textcolor{BrickRed}{#2}%
    \fi%
    }]}

\usepackage{fancyhdr}
\author{\small\scshape S\lowercase{teven} D\lowercase{uplij} \lowercase{and} R\lowercase{aimund} V\lowercase{ogl}}

\address{% \small \scshape
Center for Information Technology,
Universit\"at M\"unster,
R\"ontgenstrasse 7-13\\
D-48149 M\"unster,
Deutschland}
\email{\small \sf douplii@uni-muenster.de;
sduplij@gmail.com;
https://ivv5hpp.uni-muenster.de/u/douplii}
\title{\large\bfseries\scshape
P\lowercase{olyander visualization of quantum walks}}

\date{\textit{of start} September 22, 2023. \textit{Date}:
\textit{of completion}
October 20, 2023.
\newline
\mbox{}\hskip 1.16em
\textit{Total}:
25
references.%
}

\renewcommand{\refname}{\textsc{References}}

\let\origsection\section
\renewcommand{\section}[1]{\sectionmark{#1}\origsection{#1}}
\let\origsubsection\subsection
\renewcommand{\subsection}[1]{\subsectionmark{#1}\origsubsection{#1}}

\makeatletter
\renewenvironment{thebibliography}[1]{%
  \@xp\origsection\@xp*\@xp{\refname}%
  \normalfont\footnotesize\labelsep .9em\relax
  \renewcommand\theenumiv{\arabic{enumiv}}\let\p@enumiv\@empty
  \vspace*{-5pt}% NEW
  \list{\@biblabel{\theenumiv}}{\settowidth\labelwidth{\@biblabel{#1}}%
    \leftmargin\labelwidth \advance\leftmargin\labelsep
    \usecounter{enumiv}}%
  \sloppy \clubpenalty\@M \widowpenalty\clubpenalty
  \sfcode`\.=\@m
}{%
  \def\@noitemerr{\@latex@warning{Empty `thebibliography' environment}}%
  \endlist
}
\makeatother

\subjclass[2010]{65T50, 81P40, 81P45, 81P68}
\keywords{walker, coin, triander, polyander, visualization, Hadamard quantum walk, Fourier transform}

\begin{document}
\mbox{}
\vspace{1cm}

%\vspace{1.5cm}
\mbox{}
%\vskip 1.5cm
\begin{abstract}
%\noindent
%TCIDATA{OutputFilter=latex2.dll}
%TCIDATA{Version=5.50.0.2953}
%TCIDATA{LaTeXparent=0,0,example.TEX}

\noindent We investigate quantum walks which play an important role in the modelling of many phenomena.  The detailed and thorough description is given to the discrete quantum walks on a line, where the total quantum state consists of quantum states of the walker and the coin. In addition to the standard walker probability distribution, we introduce the coin probability distribution which gives more complete quantum walk description and novel visualization in terms of the so called polyanders (analogs of trianders in DNA visualization). The methods of final states computation and the Fourier transform are presented for the Hadamard quantum walk.

\end{abstract}
\maketitle

%\vspace{-1cm}
\thispagestyle{empty}

%\mbox{}

%\newpage
\mbox{}
\vspace{1.5cm}
%\begin{small}
\tableofcontents
%\end{small}%
\newpage

\pagestyle{fancy}

\addtolength{\footskip}{15pt}

\renewcommand{\sectionmark}[1]{%
\markboth{
%\textmd{\  \thesection.}
{ \scshape #1}}{}}

\renewcommand{\subsectionmark}[1]{%
\markright{
\mbox{\;}\\[5pt]
\textmd{#1}}{}}

\fancyhead{}
\fancyhead[EL,OR]{\leftmark}
\fancyhead[ER,OL]{\rightmark}
\fancyfoot[C]{\scshape -- \textcolor{BrickRed}{\thepage} --}

\renewcommand\headrulewidth{0.5pt}
\fancypagestyle {plain1}{ %
\fancyhf{}
\renewcommand {\headrulewidth }{0pt}
\renewcommand {\footrulewidth }{0pt}
}

\fancypagestyle{plain}{ %
\fancyhf{}
\fancyhead[C]{\scshape S\lowercase{teven} D\lowercase{uplij} \hskip 0.7cm \MakeUppercase{Polyadic Hopf algebras and quantum groups}}
\fancyfoot[C]{\scshape - \thepage  -}
\renewcommand {\headrulewidth }{0pt}
\renewcommand {\footrulewidth }{0pt}
}

\fancypagestyle{fancyref}{ %
\fancyhf{} % remove everything
\fancyhead[C]{\scshape R\lowercase{eferences} }
\fancyfoot[C]{\scshape -- \textcolor{BrickRed}{\thepage} --}
\renewcommand {\headrulewidth }{0.5pt}
\renewcommand {\footrulewidth }{0pt}
}

\fancypagestyle{emptyf}{
\fancyhead{}
\fancyfoot[C]{\scshape -- \textcolor{BrickRed}{\thepage} --}
\renewcommand{\headrulewidth}{0pt}
}
\mbox{}
%\vskip 3.5cm
\thispagestyle{emptyf}
%TCIDATA{OutputFilter=latex2.dll}
%TCIDATA{Version=5.50.0.2953}
%TCIDATA{LaTeXparent=0,0,example.TEX}

\section{\textsc{Introduction}}

Quantum walks are the quantum counterpart of the classical random walks
playing important role in the modelling of many phenomena, for instance
information spreding in complex networks \cite{noh/rie}, optimal search
strategies \cite{lv/cao/cohen}, genetic sequence location \cite{eng/sac/tra},
and chemical reactions \cite{gil77}. The term \textquotedblleft quantum
walks\textquotedblright\ was introduced in \cite{aha/dav/zag}, but the idea to
incorporate quantum effects to stochastic calculus appeared in \cite{ich/noz},
that is the coherence effects in evolution of Brownian quantum particle were
first considered in \cite{schw61}. Then the quantum analogies of classical
random walks in discrete time and space were investigated in \cite{god/fuj},
the quantum cellular automata were introduced in \cite{gro/zei} which appeared
to be equivalent to the construction of \cite{aha/dav/zag}, which can be
considered as one particle sector of the former, for a review, see
\cite{arr2019} and more general \cite{ven2012}. The connections between
correlated classical random walks and quantum walks were given in
\cite{kon2009} using matrix methods.

There are two models of quantum walks:

\begin{enumerate}
\item Discrete quantum walks consists of two systems, called a walker and a
coin, and the evolution unitary operator acts on them in discrete time steps.

\item Continuous quantum walks consists of one quantum system called walker
which \textquotedblleft walks\textquotedblright\ without time restrictions,
which is described by the evolution operator (Hamiltonian) and the
Schr\"{o}dinger equation \cite{chi/far/gut}.
\end{enumerate}

The general topology in both cases can be described by discrete graphs.

\section{\textsc{Discrete quantum walks}}

In the case of discrete quantum walks on a line the total quantum state
consists of quantum states of the walker and the coin, that is the total
Hilbert state $\mathfrak{H}_{tot}$ becomes the direct product%
\begin{equation}
\mathfrak{H}_{tot}=\mathfrak{H}_{coin}\otimes\mathfrak{H}_{walk}. \label{qw-h}%
\end{equation}

The \textquotedblleft position\textquotedblright\ of the walker is described
by the vector from the computational basis of the walker Hilbert space
$\left\vert \psi_{walk}\right\rangle \in\mathfrak{H}_{walk}$ which is
infinite-dimensional and countable, such that the walker state $\left\vert
\psi_{walk}\right\rangle $ is the quantum superposition%
\begin{equation}
\left\vert \psi_{walk}\right\rangle =\sum_{\ell\in\mathbb{Z}}w_{\ell
}\left\vert \ell\right\rangle _{w},\ \ \ \ \ \ \ \ \ \sum_{\ell\in\mathbb{Z}%
}w_{\ell}^{2}=1,\ \ \ w_{\ell}\in\mathbb{C}. \label{qw-fw}%
\end{equation}

In distinction of the classical coin which can be in two states, the quantum
$s$-state coin can be not only in $s$ canonical basis states $\left\vert
\mathbf{0}\right\rangle _{c},\left\vert \mathbf{1}\right\rangle _{c}%
,\ldots,\left\vert \mathbf{s-1}\right\rangle _{c}$, but also in their quantum
superposition%
\begin{equation}
\left\vert \mathbf{\psi}_{coin}\right\rangle =\sum_{j=0}^{s-1}c_{j}\left\vert
\mathbf{j}\right\rangle _{c},\ \ \ \ \ \ \ \ \ \sum_{j=0}^{s-1}c_{j}%
^{2}=1,\ \ \ c_{j}\in\mathbb{C}. \label{qw-fc}%
\end{equation}

Usually, to be closer to the classical case, one puts $s=2$. The total state
of the quantum walk is given by%
\begin{equation}
\left\vert \Psi_{tot}\right\rangle =\left\vert \mathbf{\psi}_{coin}%
\right\rangle \otimes\left\vert \psi_{walk}\right\rangle , \label{qw-ftot}%
\end{equation}
and the initial total state, if to take $\left\vert \psi_{walk}\right\rangle
_{initial}=\left\vert 0\right\rangle _{w}$, becomes%
\begin{equation}
\left\vert \Psi_{tot}\right\rangle _{initial}=\left\vert \mathbf{\psi}%
_{coin}\right\rangle _{initial}\otimes\left\vert 0\right\rangle _{w}.
\label{qw-in}%
\end{equation}

In general, the total state can be written as%
\begin{align}
\left\vert \Psi_{tot}\right\rangle  &  =\sum_{\ell\in\mathbb{Z}}\left(
\varphi_{0,\ell}\left\vert \mathbf{0}\right\rangle _{c}\otimes\left\vert
\ell\right\rangle _{w}+\varphi_{1,\ell}\left\vert \mathbf{1}\right\rangle
_{c}\otimes\left\vert \ell\right\rangle _{w}\right)  ,\label{qw-ft}\\
\sum_{\ell\in\mathbb{Z}}\left(  \left\vert \varphi_{0,\ell}\right\vert
^{2}+\left\vert \varphi_{1,\ell}\right\vert ^{2}\right)   &  =1\ \ \ \varphi
_{0,\ell},\varphi_{1,\ell}\in\mathbb{C}. \label{qw-t}%
\end{align}

It follows from (\ref{qw-fw})--(\ref{qw-fc}) that%
\begin{equation}
\varphi_{j,\ell}=c_{j}w_{\ell},\ \ \ \ell\in\mathbb{Z},\ \ \ j=0,1,
\label{qw-uj}%
\end{equation}
and so the normalization condition (\ref{qw-t}) reduces one parameter from the
set of ones describing the total state (\ref{qw-ft}).

By analogy with the classical random walk, we need one operator to move the
walker on the line and one operator to play the same role as the coin toss. As
opposite to the classic case, where such an operator is represented by a
stochastic matrix, in the case of the quantum walk evolution there is no room
for randomness before measurement and it is represented by an unitary matrix
which acts as an internal rotation in the internal state space. The goal of
the coin operator is to render the coin state in a superposition, while the
randomness is introduced by making a measurement on the system after both
evolution operators have been applied to the total quantum system for many times.

Thus, the evolution of a quantum walk is driven by the special composite
action of two unitary operators: 1) the first one, shift operator $\mathbf{S}$
acting in combined total position-coin space $\mathfrak{H}_{tot}$; 2) the
other one is the coin operator $\mathbf{C}$ acting in the coin space
$\mathfrak{H}_{coin}$. In this way the total evolution is described by the
unitary operator $\mathbf{U}$ defined by the main formula of the coined
quantum walk concept%
\begin{align}
\mathbf{U}  &  =\mathbf{S}\circ\left(  \mathbf{C}\otimes\mathbf{I}_{w}\right)
,\label{qw-u}\\
\mathbf{S}  &  :\mathfrak{H}_{coin}\otimes\mathfrak{H}_{walk}\rightarrow
\mathfrak{H}_{coin}\otimes\mathfrak{H}_{walk},\ \ \mathbf{C}:\mathfrak{H}%
_{coin}\rightarrow\mathfrak{H}_{coin},\ \ \mathbf{U}:\mathfrak{H}%
_{tot}\rightarrow\mathfrak{H}_{tot},
\end{align}
where $\mathbf{I}_{w}\in\mathfrak{H}_{walk}$ is the unity of the walker space
$\mathfrak{H}_{walk}$.

If we consider the two-state coin $s=2$ (\ref{qw-ft}), then the operator
$\mathbf{S}$ should act on the total quantum state (\ref{qw-ftot}) by shifts
which are dependent from the coin state%
\begin{align}
\mathbf{S\circ}\left(  \left\vert \mathbf{0}\right\rangle _{c}\otimes
\left\vert \ell\right\rangle _{w}\right)   &  =\left\vert \mathbf{0}%
\right\rangle _{c}\otimes\left\vert \ell+1\right\rangle _{w},\label{qw-s1}\\
\mathbf{S\circ}\left(  \left\vert \mathbf{1}\right\rangle _{c}\otimes
\left\vert \ell\right\rangle _{w}\right)   &  =\left\vert \mathbf{1}%
\right\rangle _{c}\otimes\left\vert \ell-1\right\rangle _{w}. \label{qw-s2}%
\end{align}

This can be written in the unified form%
\begin{equation}
\mathbf{S\circ}\left(  \left\vert \mathbf{j}\right\rangle _{c}\otimes
\left\vert \ell\right\rangle _{w}\right)  =\left\vert \mathbf{j}\right\rangle
_{c}\otimes\left\vert \ell+\left(  -1\right)  ^{j}\right\rangle _{w},
\label{qw-sj}%
\end{equation}
that is we have the shift operator depennds on the coin state $\mathbf{S}%
=\mathbf{S}_{j}$. Therefore, in the computational basis $\mathbf{S}$ can be
presentes using two projections in $\mathfrak{H}_{c}$ as (the outer product
representation)%
\begin{equation}
\mathbf{S}=\left\vert \mathbf{0}\right\rangle _{c}\left\langle \mathbf{0}%
\right\vert _{c}\otimes\sum_{\ell\in\mathbb{Z}}\left\vert \ell+1\right\rangle
_{w}\left\langle \ell\right\vert _{w}+\left\vert \mathbf{1}\right\rangle
_{c}\left\langle \mathbf{1}\right\vert _{c}\otimes\sum_{\ell\in\mathbb{Z}%
}\left\vert \ell-1\right\rangle _{w}\left\langle \ell\right\vert _{w},
\label{qw-s}%
\end{equation}
which satisfies the needed shifting properties in the walker space
(\ref{qw-s1})--(\ref{qw-s2}).

The coin operator $\mathbf{C}$ is an arbitrary element of the unitary group
$\mathcal{U}\left(  s\right)  $, and for the two-state coin $s=2$, and it can
be represented by the $4$ real parameter $2\times2$ complex matrix
$\mathbf{\hat{C}}$ of the form%
\begin{equation}
\mathbf{\hat{C}}=\mathbf{\hat{C}}_{\alpha,\beta,\gamma,\theta}=\left(
\begin{array}
[c]{cc}%
a & b\\
c & d
\end{array}
\right)  =e^{i\gamma}\left(
\begin{array}
[c]{cc}%
e^{i\alpha}\cos\theta & e^{i\beta}\sin\theta\\
-e^{-i\beta}\sin\theta & e^{-i\alpha}\cos\theta
\end{array}
\right)  ,\ \ a,b,c,d\in\mathbb{C},\ \ \alpha,\beta,\gamma,\theta\in
\mathbb{R}. \label{qw-c}%
\end{equation}

In the most cases, for quantum walks with two-state coin the Hadamard operator
is widely used%
\begin{equation}
\mathbf{C}_{H}=\frac{1}{\sqrt{2}}\left(  \left\vert \mathbf{0}\right\rangle
_{c}\left\langle \mathbf{0}\right\vert _{c}+\left\vert \mathbf{0}\right\rangle
_{c}\left\langle \mathbf{1}\right\vert _{c}+\left\vert \mathbf{1}\right\rangle
_{c}\left\langle \mathbf{0}\right\vert _{c}-\left\vert \mathbf{1}\right\rangle
_{c}\left\langle \mathbf{1}\right\vert _{c}\right)  , \label{qw-ch}%
\end{equation}
or in the matrix representation (\ref{qw-c})%
\begin{equation}
\mathbf{\hat{C}}_{H}=\mathbf{\hat{C}}_{\alpha=\frac{\pi}{2},\beta=\frac{\pi
}{2},\gamma=\frac{\pi}{2},\theta=\frac{\pi}{4}}=\frac{1}{\sqrt{2}}\left(
\begin{array}
[c]{cc}%
1 & 1\\
1 & -1
\end{array}
\right)  . \label{qw-ch1}%
\end{equation}

The evolution of the total state (\ref{qw-ftot}) during the descrete time
($=t$) quantum walk after $t$ steps $\left\vert \Psi_{tot}\left(  t\right)
\right\rangle $ is given by the application of the unitary operator
(\ref{qw-u}) $t$ times in the following way%
\begin{equation}
\left\vert \Psi_{tot}\left(  t\right)  \right\rangle =\mathbf{U}^{t}\left\vert
\Psi_{tot}\left(  0\right)  \right\rangle , \label{qw-ut}%
\end{equation}
where $\left\vert \Psi_{tot}\left(  0\right)  \right\rangle =\left\vert
\Psi_{tot}\right\rangle _{initial}$ (\ref{qw-in}).

\begin{example}
\label{qw-exam1}Using (\ref{qw-u}) and (\ref{qw-ch}) we can get the first $3$
steps for the Hadamard quantum walk with the two-state coin as%
\begin{align}
\left\vert \Psi_{tot}\left(  1\right)  \right\rangle  &  =\frac{1}{\sqrt{2}%
}\left\vert \mathbf{0}\right\rangle _{c}\otimes\left\vert 1\right\rangle
_{w}\ \ \ +\ \ \ \frac{1}{\sqrt{2}}\left\vert \mathbf{1}\right\rangle
_{c}\otimes\left\vert -1\right\rangle _{w},\label{qw-f1}\\
\left\vert \Psi_{tot}\left(  2\right)  \right\rangle  &  =-\frac{1}%
{2}\left\vert \mathbf{1}\right\rangle _{c}\otimes\left\vert -2\right\rangle
_{w}\ \ \ +\ \ \ \frac{1}{2}\left(  \left\vert \mathbf{0}\right\rangle
_{c}+\left\vert \mathbf{1}\right\rangle _{c}\right)  \otimes\left\vert
0\right\rangle _{w}\ \ \ +\ \ \ \frac{1}{2}\left\vert \mathbf{0}\right\rangle
_{c}\otimes\left\vert 2\right\rangle _{w}\label{qw-f2}\\
&  =\frac{1}{2}\left\vert \mathbf{0}\right\rangle _{c}\otimes\left(
\left\vert 0\right\rangle _{w}+\left\vert 2\right\rangle _{w}\right)
\ \ \ +\ \ \ \frac{1}{2}\left\vert \mathbf{1}\right\rangle _{c}\otimes\left(
\left\vert 0\right\rangle _{w}-\left\vert -2\right\rangle _{w}\right)
,\label{qw-f3}\\
\left\vert \Psi_{tot}\left(  3\right)  \right\rangle  &  =\frac{1}{2\sqrt{2}%
}\left\vert \mathbf{1}\right\rangle _{c}\otimes\left\vert -3\right\rangle
_{w}-\frac{1}{2\sqrt{2}}\left\vert \mathbf{0}\right\rangle _{c}\otimes
\left\vert -1\right\rangle _{w}+\frac{1}{2\sqrt{2}}\left(  2\left\vert
\mathbf{0}\right\rangle _{c}+\left\vert \mathbf{1}\right\rangle _{c}\right)
\otimes\left\vert 1\right\rangle _{w}+\frac{1}{2\sqrt{2}}\left\vert
\mathbf{0}\right\rangle _{c}\otimes\left\vert 3\right\rangle _{w}%
\label{qw-f4}\\
&  =\frac{1}{2\sqrt{2}}\left\vert \mathbf{0}\right\rangle _{c}\otimes\left(
-\left\vert -1\right\rangle _{w}+2\left\vert 1\right\rangle _{w}+\left\vert
3\right\rangle _{w}\right)  \ \ \ +\ \ \ \frac{1}{2\sqrt{2}}\left\vert
\mathbf{1}\right\rangle _{c}\otimes\left(  \left\vert 1\right\rangle
_{w}+\left\vert -3\right\rangle _{w}\right)  . \label{qw-f5}%
\end{align}

\end{example}

If the final state at the time $t$ is known $\Psi_{tot}\left(  t\right)  $,
the standard way to describe the quantum walk is the partial measurement of
the walker state probabilities (see, e.g. \cite{portugal}).

However, now we have the tensor product of two spaces (\ref{qw-h}), therefore
to have the complete description of the quantum walk we propose to consider
the partial measurement of the ($s$-) coin state probabilities as well.

Let the total state at the time $t$ (\ref{qw-ut}) has the general form (see
(\ref{qw-ft})--(\ref{qw-uj}))%
\begin{align}
\left\vert \Psi_{tot}\left(  t\right)  \right\rangle  &  =\sum_{\ell
\in\mathbb{Z}}\sum_{j=0}^{s-1}\varphi_{j,\ell}\left(  t\right)  \left\vert
\mathbf{j}\right\rangle _{c}\otimes\left\vert \ell\right\rangle _{w}%
,\label{qw-fj}\\
\sum_{\ell\in\mathbb{Z}}\left\vert \varphi_{j,\ell}\left(  t\right)
\right\vert ^{2}  &  =1\ \ \ \varphi_{j,\ell}\in\mathbb{C}.
\end{align}

We denote the \textquotedblleft doubly partial\textquotedblright\ probability
of the state $\left\vert \mathbf{j}\right\rangle _{c}\otimes\left\vert
\ell\right\rangle _{w}$ at the time $t$ by%
\begin{equation}
p_{j,\ell}\left(  t\right)  =\left\vert \varphi_{j,\ell}\left(  t\right)
\right\vert ^{2},\ \ \ \ \ \sum_{\ell\in\mathbb{Z}}\sum_{j=0}^{s-1}p_{j,\ell
}\left(  t\right)  =1.
\end{equation}

Now we propose to characterize the quantum walk by two partial probability distributions:

\begin{enumerate}
\item The walker probability distribution%
\begin{align}
p_{\ell}^{walk}\left(  t\right)   &  =\sum_{j=0}^{s-1}\left\vert
\varphi_{j,\ell}\left(  t\right)  \right\vert ^{2},\label{qw-pw}\\
\sum_{\ell\in\mathbb{Z}}p_{\ell}^{walk}\left(  t\right)   &  =1.
\label{qw-pw1}%
\end{align}

\item The coin probability distribution%
\begin{align}
p_{j}^{coin}\left(  t\right)   &  =\sum_{\ell\in\mathbb{Z}}\left\vert
\varphi_{j,\ell}\left(  t\right)  \right\vert ^{2},\label{qw-pc}\\
\sum_{j=0}^{s-1}p_{j}\left(  t\right)   &  =1. \label{qw-pc1}%
\end{align}

\end{enumerate}

In the standard approach \cite{portugal}, only the first (walker) distribution
(\ref{qw-pw}) is usually considered: the time is fixed by $t=t_{0}$, and the
graph $\left\{  \ell,p_{\ell}^{walk}\left(  t_{0}\right)  \right\}  $ is
plotted. Nevertheless, the coin probability distribution (\ref{qw-pc}) gives
additional and information about the quantum walk. To observe the difference
between (\ref{qw-pw}) and (\ref{qw-pc}) concretely, we continue the
\textit{Example} \ref{qw-exam1} in very details.

\begin{example}
[\textit{Example} \ref{qw-exam1} continued]Here we compute the walker and coin
probabilities (\ref{qw-pw}) and (\ref{qw-pc}) for three steps $t=1,2,3$ of the
Hadamard walk $\Psi_{tot}\left(  t\right)  $ in (\ref{qw-f1})--(\ref{qw-f5}%
).The formulas (\ref{qw-f1}), (\ref{qw-f2}) and (\ref{qw-f4}) are convenient
to use for the walker probabilities, and the formulas (\ref{qw-f1}),
(\ref{qw-f3}) and (\ref{qw-f5}) can be used for the coin probabilities. We
derive the walker probabilities $p_{\ell}^{walk}\left(  t\right)  $ from
(\ref{qw-f1})%
\begin{align}
&  p_{\ell=1}^{walk}\left(  t=1\right)  =p_{\ell=\left\vert 1\right\rangle
_{w}}^{walk}\left(  t=1\right)  =\left(  \frac{1}{\sqrt{2}}\right)  ^{2}%
=\frac{1}{2},\\
&  p_{\ell=-1}^{walk}\left(  t=1\right)  =p_{\ell=\left\vert -1\right\rangle
_{w}}^{walk}\left(  t=1\right)  =\left(  \frac{1}{\sqrt{2}}\right)  ^{2}%
=\frac{1}{2},
\end{align}
and from (\ref{qw-f2}) we obtain the symmetric distribution%

\begin{align}
&  p_{\ell=-2}^{walk}\left(  t=2\right)  =p_{\ell=\left\vert -2\right\rangle
_{w}}^{walk}\left(  t=2\right)  =\left(  \frac{1}{2}\right)  ^{2}=\frac{1}%
{4},\\
&  p_{\ell=0}^{walk}\left(  t=2\right)  =p_{\ell=\left\vert 0\right\rangle
_{w}}^{walk}\left(  t=2\right)  =\left(  \frac{1}{2}\right)  ^{2}+\left(
\frac{1}{2}\right)  ^{2}=\frac{1}{2},\\
&  p_{\ell=2}^{walk}\left(  t=2\right)  =p_{\ell=\left\vert 2\right\rangle
_{w}}^{walk}\left(  t=2\right)  =\left(  \frac{1}{2}\right)  ^{2}=\frac{1}{4}.
\end{align}

The probability distribution $p_{\ell}^{walk}\left(  t\right)  $ for the third
step $t=3$ is nonsymmetric (\ref{qw-f4})%
\begin{align}
&  p_{\ell=-3}^{walk}\left(  t=3\right)  =p_{\ell=\left\vert -3\right\rangle
_{w}}^{walk}\left(  t=3\right)  =\left(  \frac{1}{2\sqrt{2}}\right)
^{2}=\frac{1}{8},\\
&  p_{\ell=-1}^{walk}\left(  t=3\right)  =p_{\ell=\left\vert -1\right\rangle
_{w}}^{walk}\left(  t=3\right)  =\left(  -\frac{1}{2\sqrt{2}}\right)
^{2}=\frac{1}{8},\\
&  p_{\ell=1}^{walk}\left(  t=3\right)  =p_{\ell=\left\vert 1\right\rangle
_{w}}^{walk}\left(  t=3\right)  =\left(  2\frac{1}{2\sqrt{2}}\right)
^{2}+\left(  \frac{1}{2\sqrt{2}}\right)  ^{2}=\frac{5}{8},\\
&  p_{\ell=3}^{walk}\left(  t=3\right)  =p_{\ell=\left\vert 3\right\rangle
_{w}}^{walk}\left(  t=3\right)  =\left(  \frac{1}{2\sqrt{2}}\right)
^{2}=\frac{1}{8},
\end{align}
as well as for further steps (times) $t>3$.

For the coin probabilities $p_{\ell}^{coin}\left(  t\right)  $ we have from
(\ref{qw-f1})%
\begin{align}
&  p_{j=0}^{coin}\left(  t=1\right)  =p_{j=\left\vert \mathbf{0}\right\rangle
_{c}}^{coin}\left(  t=1\right)  =\left(  \frac{1}{\sqrt{2}}\right)  ^{2}%
=\frac{1}{2},\\
&  p_{j=1}^{coin}\left(  t=1\right)  =p_{j=\left\vert \mathbf{1}\right\rangle
_{c}}^{coin}\left(  t=1\right)  =\left(  \frac{1}{\sqrt{2}}\right)  ^{2}%
=\frac{1}{2},
\end{align}
and from (\ref{qw-f3}) we have for the second step $t=2$ the symmetric
distribution%
\begin{align}
&  p_{j=0}^{coin}\left(  t=2\right)  =p_{j=\left\vert \mathbf{0}\right\rangle
_{c}}^{coin}\left(  t=2\right)  =\left(  \left(  \frac{1}{2}\right)
^{2}+\left(  \frac{1}{2}\right)  ^{2}\right)  =\frac{1}{2},\\
&  p_{j=1}^{coin}\left(  t=2\right)  =p_{j=\left\vert \mathbf{1}\right\rangle
_{c}}^{coin}\left(  t=2\right)  =\left(  \left(  \frac{1}{2}\right)
^{2}+\left(  -\frac{1}{2}\right)  ^{2}\right)  =\frac{1}{2},
\end{align}

The probability distribution $p_{j}^{coin}\left(  t\right)  $ for the third
step $t=3$ is also nonsymmetric as $p_{\ell}^{walk}\left(  t=3\right)  $, so
from (\ref{qw-f5}) we get%
\begin{align}
&  p_{j=0}^{coin}\left(  t=3\right)  =p_{j=\left\vert \mathbf{0}\right\rangle
_{c}}^{coin}\left(  t=3\right)  =\left(  \left(  -\frac{1}{2\sqrt{2}}\right)
^{2}+\left(  2\frac{1}{2\sqrt{2}}\right)  ^{2}+\left(  \frac{1}{2\sqrt{2}%
}\right)  ^{2}\right)  =\frac{3}{4},\\
&  p_{j=1}^{coin}\left(  t=3\right)  =p_{j=\left\vert \mathbf{1}\right\rangle
_{c}}^{coin}\left(  t=3\right)  =\left(  \left(  \frac{1}{2\sqrt{2}}\right)
^{2}+\left(  2\frac{1}{2\sqrt{2}}\right)  ^{2}\right)  =\frac{1}{4},
\end{align}
and in the similar way for further steps (discrete times) $t>3$.

As it should be, both the above walker and coin probability distributions are
correctly normalized satisfying (\ref{qw-pw1}) and (\ref{qw-pc1}) at each
discrete time $t$.
\end{example}

\section{\label{subsec-poly}\textsc{Polyander visualization of quantum walks}}

The coin probability distribution $p_{j}^{coin}\left(  t\right)  $ introducted
in (\ref{qw-pc}), from the first glance, can be also characterized at the
fixed time $t=t_{0}$ by the graph $\left\{  j,p_{j}^{coin}\left(
t_{0}\right)  \right\}  $, as the walker probability distribution $p_{\ell
}^{walk}\left(  t_{0}\right)  $. However, because the coin has a specific
\textquotedblleft physical\textquotedblright\ sense, we propose here another
way of the quantum walk description, which has an origin from genome
landscapes \cite{azb73,azb95,lob96}, one-dimensional DNA walks \cite{ceb/dud}
and trianders \cite{dup/dup2005}.

\begin{criterion}
We can consider the time evolution of the probability for the concrete quantum
state, when we provide the corresponding measurements in the coin or walker
subspaces. That is, we fix the states $\ell=\ell_{0}$ or $j=j_{0}$ and
introduce the following time evolution graphs $\left\{  t,p_{\ell=\ell_{0}%
}^{walk}\left(  t\right)  \right\}  $ or $\left\{  t,p_{j=j_{0}}^{coin}\left(
t\right)  \right\}  $.
\end{criterion}

\begin{definition}
The polyander visualization of a quantum walk is its description by the time
evolution graphs $\left\{  t,p_{\ell}^{walk}\left(  t\right)  \right\}  $ or
$\left\{  t,p_{j}^{coin}\left(  t\right)  \right\}  $. Each line of the graph
describing the probability evolution of the fixed quantum state $\ell=\ell
_{0}$ for $\left\vert \ell_{0}\right\rangle _{w}$ or $j=j_{0}$ for $\left\vert
\mathbf{j}\right\rangle _{c}$ is called a leg of the polyander.
\end{definition}

It is obvious, that the walker polyander has finitely increasing number of
legs and corresponding quantum states, while the $s$-side coin polyander has
exactly $s$ legs.

For the \textit{Example }\ref{qw-exam1} we obtain

\begin{example}
[\textit{Example }\ref{qw-exam1} continued]\label{qw-exam1w}The walker
polyander $p_{\ell}^{walk}\left(  t\right)  $ in the time range $1\leq t\leq3$
has $7$ legs (quantum states) $-3\leq\ell\leq3$, which have the following
probability evolutions%
\begin{equation}%
\begin{tabular}
[c]{|c||c|c|c|}\hline
$\left\vert \ell\right\rangle $-leg$\setminus$time $t$ & $1$ & $2$ &
$3$\\\hline\hline
$\left\vert -3\right\rangle _{w}$ & $0$ & $0$ & $\frac{1}{8}$\\\hline
$\left\vert -2\right\rangle _{w}$ & $0$ & $\frac{1}{4}$ & $0$\\\hline
$\left\vert -1\right\rangle _{w}$ & $\frac{1}{2}$ & $0$ & $\frac{1}{8}%
$\\\hline
$\left\vert 0\right\rangle _{w}$ & $0$ & $\frac{1}{2}$ & $0$\\\hline
$\left\vert 1\right\rangle _{w}$ & $\frac{1}{2}$ & $0$ & $\frac{5}{8}$\\\hline
$\left\vert 2\right\rangle _{w}$ & $0$ & $\frac{1}{4}$ & $0$\\\hline
$\left\vert 3\right\rangle _{w}$ & $0$ & $0$ & $\frac{1}{8}$\\\hline
\end{tabular}
\end{equation}

The coin polyander $p_{j}^{coin}\left(  t\right)  $ in the time range $1\leq
t\leq3$ has $2$ legs (quantum states), $j=0,1$, which have the following
probability evolutions%
\begin{equation}%
\begin{tabular}
[c]{|c||c|c|c|}\hline
$\left\vert \mathbf{j}\right\rangle $-leg$\setminus$time $t$ & $1$ & $2$ &
$3$\\\hline\hline
$\left\vert \mathbf{0}\right\rangle _{c}$ & $\frac{1}{2}$ & $\frac{1}{2}$ &
$\frac{3}{4}$\\\hline
$\left\vert \mathbf{1}\right\rangle _{c}$ & $\frac{1}{2}$ & $\frac{1}{2}$ &
$\frac{1}{4}$\\\hline
\end{tabular}
\ \ \label{qw-tc}%
\end{equation}

Each leg can be presented as a horizontal strip of the width $1$ on which the
points corresponding to the probabilities $0\leq p\left(  t\right)  \leq1$ at
times $t=1,2,3\ldots$ are indicated. Then the probability behaviour of each
quantum state can be visually seen and mutually compared in the same time points.
\end{example}

For the coin polyader it is important to consider the probability differences,
because of the following

\begin{definition}
The total quantum state is called trivial at the time $t=t_{triv}$, if all the
$s$-side coin states have equal probabilities $p_{j}^{coin}\left(
t_{triv}\right)  =\frac{1}{s}$, $j=0,1,\ldots,s-1$, $s\geq2$.
\end{definition}

\begin{definition}
The quantum walk is called trivial, if the $s$-side coin states are trivial at
all times.
\end{definition}

In the case of the standard coin $s=2$, the triviality means that the
measurements of both sides give the same probability at the $t=t_{triv}$.
Therefore, to describe triviality in detail, we should introduce the
differences and search for nonzero ones.

\begin{definition}
The bias $s$-side coin polyander has $\left(  s-1\right)  $ legs which are
defined by%
\begin{equation}
\Delta p_{j}^{coin}\left(  t\right)  =p_{j}^{coin}\left(  t\right)
-p_{j+1}^{coin}\left(  t\right)  ,\ \ \ j=0,\ldots,s-2.
\end{equation}

\end{definition}

\begin{example}
[\textit{Example }\ref{qw-exam1} continued]\label{qw-exam1c}The $2$-side coin
bias polyander in the time range $1\leq t\leq3$ has one leg which has the
following probability evolution $\Delta p_{0}^{coin}\left(  t\right)  =\Delta
p_{j=\left\vert \mathbf{0}\right\rangle _{c}}^{coin}\left(  t\right)  -\Delta
p_{j=\left\vert \mathbf{1}\right\rangle _{c}}^{coin}\left(  t\right)  $ (see
(\ref{qw-tc}))%
\begin{equation}%
\begin{tabular}
[c]{|c||c|c|c|}\hline
$j\setminus$time $t$ & $1$ & $2$ & $3$\\\hline\hline
$0$ & $0$ & $0$ & $\frac{1}{2}$\\\hline
\end{tabular}
\end{equation}
which can be nontrivial after the time $t=3$ only.
\end{example}

In the higher times the walker and coin polyanders, as well as the bias coin
polyander will have more complated behavior, which in any case needs the
manifest form of the total quantum state (\ref{qw-ut}). In \textit{Example}
\ref{qw-exam1w} and \textit{Example \ref{qw-exam1c}}, we considered for
clarity only the time range $1\leq t\leq3$ and the $2$-side coin to show in
details, how to compute probability polyanders for finite times. The
\textquotedblleft physical sense\textquotedblright\ of the bias polyander is
in the following: its nonzero values show nontriviality evolution along the
quantum walk.

Thus, polyanders allow us to study further the \textquotedblleft fine
structure\textquotedblright\ and thorough characterization and visual
presentation of quantum walks from different vieponts.

\section{\textsc{Methods of final states computation}}

The main goal of studying the quantum walks is obtaining the analytical
expression for the final quantum state (\ref{qw-ut}) in discrete finite times
$t\in\mathbb{Z}$, and then calculating the dynamical and statistical
properties of various probability distributions and characteristics.

The main computational methods to find the total quantum state (\ref{qw-ut}) are

\begin{enumerate}
\item \textbf{The Schr\"{o}dinger approach}. Starting from an arbitrary state
of the quantum walk with a certain walker position, to provide the discrete
time Fourier transform \cite{amb/bac/nay} and obtain the closed form of total amplitudes.

\item \textbf{The combinatorial approach}. The amplitude at any discrete time
is derived as a sum of amplitudes of all paths starting from the initial state
and ending up in the final state. This can be treated as reminiscent of the
standard path integral technique.
\end{enumerate}

In \cite{car/ric/tem} it was shown that both Schr\"{o}dinger and combinatorial
approaches are equivalent. Among less known methods we can mention the
alternative description of quantum walks based on the scattering theory
\cite{fel/hil} and the analytic formulation of probability densities and
moments \cite{fus/whi/she}.

\subsection{Fourier transform and analytic solutions}

In general, the usage of the Fourier transform is the standard way of
simplification of computations by turning equations to the algebraic ones. In
its application to quantum works and analysing the evolution (\ref{qw-ut})
there two peculiarities:

\begin{enumerate}
\item The Fourier transform is applied to one subspace from the product
(\ref{qw-ftot}), that is the walker one $\mathfrak{H}_{walk}$.

\item Sometimes it is more simple to turn from transforming functions to
transform the computational basis of the walker subspace.
\end{enumerate}

Following \emph{2)} we transform the computational basis of the walker space
$\mathfrak{H}_{walk}$ as%
\begin{equation}
\left\vert \left\vert \mathsf{k}\right\rangle \right\rangle _{w}=\sum_{\ell
\in\mathbb{Z}}e^{i\mathsf{k}\ell}\left\vert \ell\right\rangle _{w}%
,\ \ \ \ell\in\mathbb{Z},\ \ \left\vert \ell\right\rangle _{w},\ \ \left\vert
\left\vert \mathsf{k}\right\rangle \right\rangle _{w}\in\mathfrak{H}_{walk},
\label{qw-kw}%
\end{equation}
where the Fourier transformed vectors $\left\vert \left\vert \mathsf{k}%
\right\rangle \right\rangle _{w}$ are denoted by the double brackets and
depend on the continuous real \textquotedblleft wave number\textquotedblright%
\ $\mathsf{k}\in\mathbb{R},$ $-\pi\leq\mathsf{k}\leq\pi$. The inverse
transformation is%
\begin{equation}
\left\vert \ell\right\rangle _{w}=\frac{1}{2\pi}\int_{-\pi}^{\pi}%
d\mathsf{k}e^{-i\mathsf{k}\ell}\left\vert \left\vert \mathsf{k}\right\rangle
\right\rangle _{w}.
\end{equation}
Let us introduce the Fourier transformation of the amplitudes $\varphi
_{j,\ell}\left(  t\right)  $ at the time $t$ from the decomposition
(\ref{qw-fj}) in the standard way by%
\begin{equation}
\Phi_{j,\mathsf{k}}\left(  t\right)  =\sum_{\ell\in\mathbb{Z}}e^{-i\mathsf{k}%
\ell}\varphi_{j,\ell}\left(  t\right)  ,\ \ \ -\pi\leq\mathsf{k}\leq\pi.
\label{qw-f}%
\end{equation}
The inverse Fourier transform becomes%
\begin{equation}
\varphi_{j,\ell}\left(  t\right)  =\frac{1}{2\pi}\int_{-\pi}^{\pi}%
d\mathsf{k}e^{i\mathsf{k}\ell}\Phi_{j,\mathsf{k}}\left(  t\right)  .
\label{qw-ff}%
\end{equation}

Then, instead of the computational basis $\left\vert \mathbf{j}\right\rangle
_{c}\otimes\left\vert \ell\right\rangle _{w}$ in (\ref{qw-fj}), using
(\ref{qw-kw}) and (\ref{qw-ff}) and cancelling exponents, we can present the
total state in the Fourier basis $\left\vert \mathbf{j}\right\rangle
_{c}\otimes\left\vert \left\vert \mathsf{k}\right\rangle \right\rangle _{w}$
as follows%
\begin{equation}
\left\vert \Psi_{tot}\left(  t\right)  \right\rangle =\frac{1}{2\pi}\sum
_{j=0}^{s-1}\int_{-\pi}^{\pi}\Phi_{j,\mathsf{k}}\left(  t\right)  \left\vert
\mathbf{j}\right\rangle _{c}\otimes\left\vert \left\vert \mathsf{k}%
\right\rangle \right\rangle _{w}. \label{qw-fto}%
\end{equation}

The action of the shift operator $\mathbf{S}$ on the Fourier basis can be
derived from (\ref{qw-sj}) and using (\ref{qw-kw}) as follows%
\begin{align}
\mathbf{S\circ}\left(  \left\vert \mathbf{j}\right\rangle _{c}\otimes
\left\vert \left\vert \mathsf{k}\right\rangle \right\rangle _{w}\right)   &
=\sum_{\ell\in\mathbb{Z}}e^{i\mathsf{k}\ell}\mathbf{S\circ}\left(  \left\vert
\mathbf{j}\right\rangle _{c}\otimes\left\vert \ell\right\rangle _{w}\right)
=\sum_{\ell\in\mathbb{Z}}e^{i\mathsf{k}\ell}\mathbf{S\circ}\left(  \left\vert
\mathbf{j}\right\rangle _{c}\otimes\left\vert \ell\right\rangle _{w}\right)
\nonumber\\
&  =\sum_{\ell\in\mathbb{Z}}e^{i\mathsf{k}\ell}\left(  \left\vert
\mathbf{j}\right\rangle _{c}\otimes\left\vert \ell+\left(  -1\right)
^{j}\right\rangle _{w}\right)  =\sum_{\ell^{\prime}\in\mathbb{Z}%
}e^{i\mathsf{k}\left(  \ell^{\prime}-\left(  -1\right)  ^{j}\right)  }\left(
\left\vert \mathbf{j}\right\rangle _{c}\otimes\left\vert \ell^{\prime
}\right\rangle _{w}\right) \nonumber\\
&  =e^{-i\mathsf{k}\left(  -1\right)  ^{j}}\sum_{\ell^{\prime}\in\mathbb{Z}%
}e^{i\mathsf{k}\ell^{\prime}}\left(  \left\vert \mathbf{j}\right\rangle
_{c}\otimes\left\vert \ell^{\prime}\right\rangle _{w}\right)  =e^{-i\mathsf{k}%
\left(  -1\right)  ^{j}}\left\vert \mathbf{j}\right\rangle _{c}\otimes
\left\vert \left\vert \mathsf{k}\right\rangle \right\rangle _{w},
\label{qw-sjk}%
\end{align}
where we used the substitution $\ell^{\prime}=\ell+\left(  -1\right)  ^{j}$
and the translation symmetry of the infinite sum.

In the case of the two-side coin $j=0,1$ and the Hadamard quantum walk
(\ref{qw-ch})--(\ref{qw-ch1}), the action of operators can be expressed in the
matrix form.

So we apply the total evolution operator $\mathbf{U}$ (\ref{qw-u}) in the
matrix form to the Fourier basis $\left\vert \mathbf{j}\right\rangle
_{c}\otimes\left\vert \left\vert \mathsf{k}\right\rangle \right\rangle _{w}$
using (\ref{qw-ch1}) to get%
\begin{align}
\mathbf{\hat{U}}\left(  \left\vert \mathbf{j}^{\prime}\right\rangle
_{c}\otimes\left\vert \left\vert \mathsf{k}\right\rangle \right\rangle
_{w}\right)   &  =\mathbf{\hat{S}}\left(  \left(  \sum_{j=0}^{1}%
\mathbf{\hat{C}}_{jj^{\prime}}\left\vert \mathbf{j}\right\rangle _{c}\right)
\otimes\left\vert \left\vert \mathsf{k}\right\rangle \right\rangle _{w}\right)
\nonumber\\
&  =\left(  \sum_{j=0}^{1}e^{-i\mathsf{k}\left(  -1\right)  ^{j}}%
\mathbf{\hat{C}}_{jj^{\prime}}\left\vert \mathbf{j}\right\rangle _{c}\right)
\otimes\left\vert \left\vert \mathsf{k}\right\rangle \right\rangle _{w}%
=\sum_{j=0}^{1}\mathbf{\bar{C}}_{jj^{\prime}}\left(  \mathsf{k}\right)
\left\vert \mathbf{j}\right\rangle _{c}\otimes\left\vert \left\vert
\mathsf{k}\right\rangle \right\rangle _{w}, \label{qw-uc}%
\end{align}
where%
\begin{equation}
\mathbf{\bar{C}}\left(  \mathsf{k}\right)  \mathbf{=}\left(
\begin{array}
[c]{cc}%
e^{-i\mathsf{k}} & 0\\
0 & e^{i\mathsf{k}}%
\end{array}
\right)  \mathbf{\hat{C}}=\frac{1}{\sqrt{2}}\left(
\begin{array}
[c]{cc}%
e^{-i\mathsf{k}} & e^{-i\mathsf{k}}\\
e^{i\mathsf{k}} & -e^{i\mathsf{k}}%
\end{array}
\right)  . \label{qw-uc1}%
\end{equation}

It follows from (\ref{qw-uc}) that diagonalization of $\mathbf{\bar{C}}\left(
\mathsf{k}\right)  $ leads to the spectral decomposition of the total operator
$\mathbf{\hat{U}}$. Indeed, if $\lambda\left(  \mathsf{k}\right)  $ is the
eigenvalue of the matrix $\mathbf{\bar{C}}\left(  \mathsf{k}\right)  $, then
it is also the eigenvalue of $\mathbf{\hat{U}}$, as it is seen from
(\ref{qw-uc}). The corresponding $\lambda\left(  \mathsf{k}\right)  $
eigenvector we denote by $\left\vert \left\vert \mathsf{v}_{\lambda\left(
\mathsf{k}\right)  }\right\rangle \right\rangle _{c}$, such that%
\begin{equation}
\mathbf{\hat{U}}\circ\left(  \left\vert \left\vert \mathsf{v}_{\lambda\left(
\mathsf{k}\right)  }\right\rangle \right\rangle _{c}\otimes\left\vert
\left\vert \mathsf{k}\right\rangle \right\rangle _{w}\right)  =\left(
\mathbf{\bar{C}}\left(  \mathsf{k}\right)  \circ\left\vert \left\vert
\mathsf{v}_{\lambda\left(  \mathsf{k}\right)  }\right\rangle \right\rangle
_{c}\right)  \otimes\left\vert \left\vert \mathsf{k}\right\rangle
\right\rangle _{w}=\lambda\left(  \mathsf{k}\right)  \left\vert \left\vert
\mathsf{v}_{\lambda\left(  \mathsf{k}\right)  }\right\rangle \right\rangle
_{c}\otimes\left\vert \left\vert \mathsf{k}\right\rangle \right\rangle _{w}.
\end{equation}

The matrix $\mathbf{\bar{C}}\left(  \mathsf{k}\right)  $ (\ref{qw-uc1}) has
two eigenvalues%
\begin{align}
\lambda_{1}\left(  \mathsf{k}\right)   &  =e^{-i\alpha\left(  \mathsf{k}%
\right)  },\ \ \ \ \ \lambda_{2}\left(  \mathsf{k}\right)  =-e^{i\alpha\left(
\mathsf{k}\right)  },\\
\alpha\left(  \mathsf{k}\right)   &  =\arcsin\left(  \frac{1}{\sqrt{2}}%
\sin\mathsf{k}\right)  ,\ \ -\frac{\pi}{2}\leq\alpha\left(  \mathsf{k}\right)
\leq\frac{\pi}{2},
\end{align}
and two corresponding normalized eigenvectors%
\begin{align}
\left\vert \left\vert \mathsf{v}_{\lambda_{1,2}\left(  \mathsf{k}\right)
}\right\rangle \right\rangle _{c}  &  =\frac{1}{\sqrt{r_{1,2}}}\left(
\begin{array}
[c]{c}%
e^{-i\mathsf{k}}\\
\pm\sqrt{2}e^{-i\alpha\left(  \mathsf{k}\right)  }-e^{-i\mathsf{k}}%
\end{array}
\right)  ,\label{qw-vl}\\
r_{1,2}  &  =2\left(  1+\cos^{2}\mathsf{k}\mp\cos\mathsf{k}\sqrt{1+\cos
^{2}\mathsf{k}}\right)  .
\end{align}

Thus, in the total evolution operator can be written in terms of eigenvalues
and eigenvectors of $\mathbf{\bar{C}}\left(  \mathsf{k}\right)  $
(\ref{qw-uc1})%
\begin{equation}
\mathbf{\hat{U}}=\frac{1}{2\pi}\int_{-\pi}^{\pi}d\mathsf{k}\left[  \left(
e^{-i\alpha\left(  \mathsf{k}\right)  }\left\vert \left\vert \mathsf{v}%
_{\lambda_{1}\left(  k\right)  }\right\rangle \right\rangle _{c}\left\langle
\left\langle \mathsf{v}_{\lambda_{1}\left(  k\right)  }\right\vert \right\vert
_{c}-e^{i\alpha\left(  \mathsf{k}\right)  }\left\vert \left\vert
\mathsf{v}_{\lambda_{2}\left(  k\right)  }\right\rangle \right\rangle
_{c}\left\langle \left\langle \mathsf{v}_{\lambda_{2}\left(  k\right)
}\right\vert \right\vert _{c}\right)  \otimes\left\vert \left\vert
\mathsf{k}\right\rangle \right\rangle _{w}\left\langle \left\langle
\mathsf{k}\right\vert \right\vert _{w}\right]  .
\end{equation}

Using orthogonality the basis eigenvectors, the power of the evolution
operator can be presented as%
\begin{equation}
\mathbf{\hat{U}}^{t}=\frac{1}{2\pi}\int_{-\pi}^{\pi}d\mathsf{k}\left[  \left(
e^{-i\alpha\left(  \mathsf{k}\right)  t}\left\vert \left\vert \mathsf{v}%
_{\lambda_{1}\left(  k\right)  }\right\rangle \right\rangle _{c}\left\langle
\left\langle \mathsf{v}_{\lambda_{1}\left(  k\right)  }\right\vert \right\vert
_{c}+\left(  -1\right)  ^{t}e^{i\alpha\left(  \mathsf{k}\right)  t}\left\vert
\left\vert \mathsf{v}_{\lambda_{2}\left(  k\right)  }\right\rangle
\right\rangle _{c}\left\langle \left\langle \mathsf{v}_{\lambda_{2}\left(
k\right)  }\right\vert \right\vert _{c}\right)  \otimes\left\vert \left\vert
\mathsf{k}\right\rangle \right\rangle _{w}\left\langle \left\langle
\mathsf{k}\right\vert \right\vert _{w}\right]  . \label{qw-ftt}%
\end{equation}

Now we can use the main quantum evolution forlmula (\ref{qw-ut}) to obtain the
total quantum state at any time from an initial quantum state (\ref{qw-in}).
For instance, if $\left\vert \Psi_{tot}\right\rangle _{initial}=\left\vert
\mathbf{0}\right\rangle _{c}\otimes\left\vert 0\right\rangle _{w}$, then using
(\ref{qw-fto}) and (\ref{qw-vl}), we derive the Fourier transformed amplitudes%
\begin{align}
\Phi_{j=0,\mathsf{k}}\left(  t\right)   &  =\frac{1}{2\sqrt{1+\cos
^{2}\mathsf{k}}}\left[  \left(  \sqrt{1+\cos^{2}\mathsf{k}}+\cos
\mathsf{k}\right)  e^{-i\alpha\left(  \mathsf{k}\right)  t}+\left(
\sqrt{1+\cos^{2}\mathsf{k}}-\cos\mathsf{k}\right)  e^{i\left(  \pi
+\alpha\left(  \mathsf{k}\right)  \right)  t}\right]  ,\nonumber\\
\Phi_{j=1,\mathsf{k}}\left(  t\right)   &  =\frac{e^{i\mathsf{k}}}%
{2\sqrt{1+\cos^{2}\mathsf{k}}}\left(  e^{-i\alpha\left(  \mathsf{k}\right)
t}-e^{i\left(  \pi+\alpha\left(  \mathsf{k}\right)  \right)  t}\right)  .
\end{align}

Then applying the reverse Fourier transform (\ref{qw-ff}) and taking into
account symmetries of integrand, we get the amplitudes in the computational
basis at the arbitrary time $t$ as%
\begin{align}
\varphi_{j=0,\ell}\left(  t\right)   &  =\left\{
\begin{array}
[c]{c}%
\dfrac{1}{2\pi}%
%TCIMACRO{\dint \limits_{-\pi}^{\pi}}%
%BeginExpansion
{\displaystyle\int\limits_{-\pi}^{\pi}}
%EndExpansion
d\mathsf{k}e^{i\left(  \mathsf{k}\ell-\alpha\left(  \mathsf{k}\right)
t\right)  }\left(  \dfrac{\cos\mathsf{k}}{\sqrt{1+\cos^{2}\mathsf{k}}%
}+1\right)  ,\ \ \ \ t+\ell=even,\\
0,\ \ \ \ \ \ \ \ \ \ \ \ \ \ \ \ \ \ \ \ \ \ \ \ \ \ \ \ \ \ \ \ \ \ \ \ \ \ \ \ \ \ \ \ \ \ \ \ \ \ \ \ \ \ \ \ \ \ \ \ \ \ \ \ \ \ \ t+\ell
=odd,
\end{array}
\right. \\
\varphi_{j=1,\ell}\left(  t\right)   &  =\left\{
\begin{array}
[c]{c}%
\dfrac{1}{2\pi}%
%TCIMACRO{\dint \limits_{-\pi}^{\pi}}%
%BeginExpansion
{\displaystyle\int\limits_{-\pi}^{\pi}}
%EndExpansion
d\mathsf{k}e^{i\left(  \mathsf{k}\ell-\alpha\left(  \mathsf{k}\right)
t+\mathsf{k}\right)  }\dfrac{1}{\sqrt{1+\cos^{2}\mathsf{k}}},\ \ \ \ t+\ell
=even,\\
0,\ \ \ \ \ \ \ \ \ \ \ \ \ \ \ \ \ \ \ \ \ \ \ \ \ \ \ \ \ \ \ \ \ \ \ \ \ \ \ \ \ \ \ \ \ \ \ \ \ \ \ \ \ \ \ \ \ t+\ell
=odd.
\end{array}
\right.
\end{align}

Finally, using the partial probability formulas (\ref{qw-pw}) and
(\ref{qw-pc}) one can plot the time evolution graphs $\left\{  t,p_{\ell
=\ell_{0}}^{walk}\left(  t\right)  \right\}  $ and $\left\{  t,p_{j=j_{0}%
}^{coin}\left(  t\right)  \right\}  $, that is to provide the polyander
visualization (see \textbf{Section \ref{subsec-poly}}).

\section{\textsc{Generalizations of descrete-time quantum walks}}

There are plenty of various generalizations of the above constructions.
Nevertheless, the main procedures remain the nearly same.

\begin{description}
\item[Coin operator] The most general form of the two-sided ($s=2$) coin
operator $\mathbf{C}$ is given by the complex matrix (\ref{qw-c}) from the
unitary group $\mathcal{U}\left(  2\right)  $. That is other than the Hadamard
matrix (\ref{qw-ch1}) can be considered, for instance the Fourier coin
\cite{portugal}.

\item[Higher dimensions] The main quantum walk equation (\ref{qw-u}) can be
extended to higher dimension the $s$-sided coin, when $\mathfrak{H}_{coin}$ is
$2s$-dimensional Hilbert space and $\mathfrak{H}_{walk}$ is the Hilbert space
corresponding to the direct product $\overset{s}{\overbrace{\mathbb{Z}%
\otimes\ldots\otimes\mathbb{Z}}}$. The common choice for $s$-sided coin is the
Grover operator described by the corresponding $2s$-dimensional matrix
$\mathbf{\hat{C}}_{Grover}$ proposed in \cite{moo/rus}.

\item[Anyonic quantum walks] To include the braiding interaction one includes
the additional Hilbert space (fusion space) $\mathfrak{H}_{fusion}$ where the
generators of the braid group act. Then the total space becomes $\mathfrak{H}%
_{tot}=\mathfrak{H}_{coin}\otimes\mathfrak{H}_{fusion}\otimes\mathfrak{H}%
_{walk}$, and the time evolution contains the additional braid operator in
some represenation \cite{leh/zat/bre}.
\end{description}

%\newpage
%\pagestyle{fancyref}
\pagestyle{emptyf}
%\mbox{}
%\vskip 0.5cm

\end{document}